\documentclass[sigconf]{acmart}

\usepackage{booktabs} % For formal tables
\usepackage{todonotes}

\setcopyright{rightsretained}
\usepackage{amsfonts}
\usepackage{amsmath} 
\usepackage{amssymb}
\usepackage{amsthm}	
\usepackage{enumerate}
\usepackage{mathrsfs}
\usepackage{booktabs}
\usepackage{epstopdf}
\usepackage[percent]{overpic}
\usepackage{bm}
\usepackage{multicol}
\usepackage{multirow}

\usepackage{mathtools}
\usepackage{empheq} 
\usepackage{cases}
\usepackage{blkarray}
\usepackage{stfloats}
\fnbelowfloat 
\usepackage{mathtools}
\usepackage{float}
 \usepackage{subfloat}
  \usepackage{subcaption}

\setcopyright{acmlicensed}
\acmConference[DAPA '19]{DAPA 2019 WSDM Workshop on Deep matching in Practical Applications}{February 15th, 2019}{Melbourne, Australia} 
\acmYear{2019}
\copyrightyear{2019}
\acmDOI{}
\acmISBN{}
\acmPrice{}

\newcommand{\Ccal}{{\mathcal C}}

\newcommand{\Lcal}{{\mathcal L}}
\newcommand{\Ical}{{\mathcal I}}
\newcommand{\Scal}{{\mathcal S}}
\newcommand{\Prob}{{\mathbb P}}

\newcommand{\bone}{{\mathbf 1}}
\newcommand{\R}{\mathbf R}

\newcommand{\Hcal}{ \mathcal H}

\acmArticle{4}
\acmPrice{15.00}

\begin{document}
\title{Learning Item-Interaction Embeddings for User Recommendations}
% \titlenote{* These authors have equally contributed to this work.}
%\subtitle{Extended Abstract}
%\subtitlenote{The full version of the author's guide is available as
%  \texttt{acmart.pdf} document}

\author{Xiaoting Zhao*, Raphael Louca*, Diane Hu, Liangjie Hong}
\thanks{*These authors have equally contributed to this work}
\affiliation{%
  \institution{Etsy, Inc}
  \city{New York}
  \country{U.S.A}}
\email{ {xzhao, rlouca, dhu, lhong}  @ etsy.com}


\begin{abstract}

Industry-scale recommendation systems have become a cornerstone of the e-commerce shopping experience. For Etsy, an online marketplace with over 50 million handmade and vintage items, users come to rely on personalized recommendations to surface relevant items from its massive inventory.  One hallmark of Etsy's shopping experience is the multitude of ways in which a user can interact with an item they are interested in: they can view it, favorite it, add it to a collection, add it to cart, purchase it, etc. We hypothesize that the different ways in which a user interacts with an item indicates different kinds of intent. Consequently, a user's recommendations should be based not only on the item from their past activity, but also the way in which they interacted with that item. In this paper, we propose a novel method for learning interaction-based item embeddings that encode the co-occurrence patterns of not only the item itself, but also the interaction type. The learned embeddings give us a convenient way of approximating the likelihood that one item-interaction pair would co-occur with another by way of a simple inner product. Because of its computational efficiency, our model lends itself naturally as a candidate set selection method, and we evaluate it as such in an industry-scale recommendation system that serves live traffic on Etsy.com. Our experiments reveal that taking interaction type into account shows promising results in improving the accuracy of modeling user shopping behavior. 

%Industry-scale recommendation systems have become a cornerstone of the e-commerce shopping experience. For Etsy, an online marketplace with over 50 million handmade and vintage items, users come to rely on personalized recommendations to surface relevant items from its massive inventory. Candidate set selection is a crucial step in this process as it efficiently prunes down the inventory to a smaller set of relevant items, making it feasible for the set to be re-ranked by a more sophisticated machine learning model. As such, algorithms for candidate set selection must be extremely computationally efficient. In previous work, candidate set selection is often based on simple heuristics or basic co-occurrence counts. In this paper, we propose a model for learning interaction-based item embeddings that encode the co-occurrence patterns of not only the item itself, but also the way in which a user interacts with them, such as by clicking, carting, or purchasing. By doing so, we can go beyond the use of explicit co-occurrence counts, and instead, generalize to learning patterns of co-occurrences between any kind of interaction types, such as click-to-carts or cart-to-purchases. We evaluate our proposed model as a candidate set selection method in an industry-scale recommendation system that serves live traffic on Etsy.com and reveal that encoding patterns of interactions shows promising results in improved modeling of user shopping behavior. 

\end{abstract}



% \ccsdesc[500]{Information systems-Content ranking}
% \ccsdesc[300]{Information systems-Web log analysis}
% \ccsdesc{Information systems-Personalization}
% \ccsdesc[100]{Information systems-Query representation;}
% \ccsdesc[100]{Information systems-Query representation;}


\maketitle

\section{Introduction}

As online shopping becomes more prevalent, and inventory grows at an exponential scale, customers have come to rely on personalized recommendation systems to understand their preferences and surface relevant items to them. For Etsy, an online, handmade marketplace with over 50 million active items, recommendation systems become even more critical to helping customers identify items of interest amidst Etsy's vast breadth of one-of-a-kind listings. 

\begin{figure}[b]
% \begin{subfigure}{0.515\textwidth}
%   \includegraphics[width=.9\linewidth]{images/target_embedding_2.png}
%   \caption{Target item}
%   \label{fig:sub1}
% \end{subfigure}\\ [0.5em]

\begin{subfigure}{0.515\textwidth}
\centering
  \includegraphics[scale=0.1]{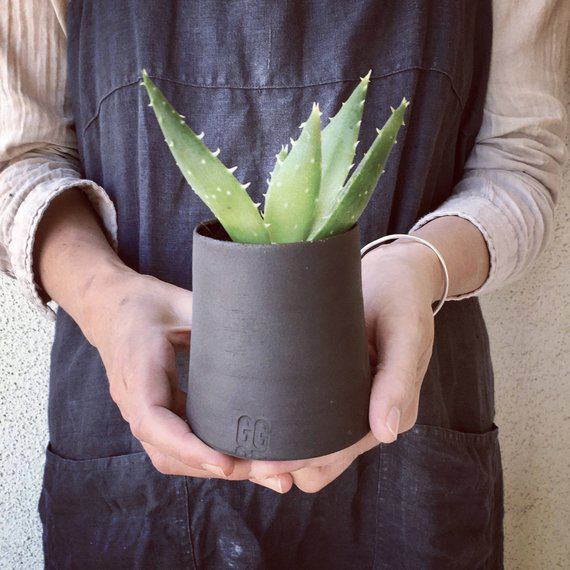}
  \caption{Target item}
  \label{fig:interaction_target}
\end{subfigure}\\ [0.5em]

%\begin{subfigure}{0.515\textwidth}
%  \includegraphics[width=.9\linewidth]{images/target_no_interaction.png}
%    \caption{Embeddings with no interactions}
 % \label{fig:sub2}
%\end{subfigure}

\begin{subfigure}{0.515\textwidth}
  \includegraphics[width=.9\linewidth]{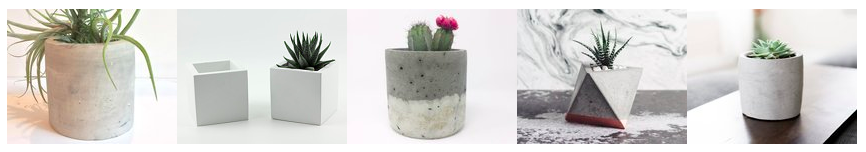}
    \caption{View-interaction embeddings }
  \label{fig:interaction_view}
\end{subfigure}

\begin{subfigure}{0.515\textwidth}
  \includegraphics[width=.9\linewidth]{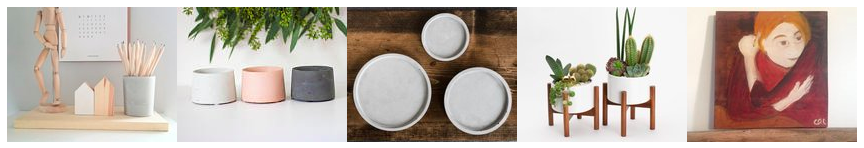}
    \caption{Cart-interaction embeddings}
  \label{fig:interaction_cart}
\end{subfigure}
\caption{Given a target item $\ell$ shown in (a), we visualize the top $5$ nearest neighbors based on $\ell$'s view-based embedding (b) and $\ell$'s cart-based embedding (c). One can see that the recommendations for a user who has viewed item $\ell$ should be different than the recommendations for a user who has already carted item $\ell$.}
\label{fig:test}
\end{figure}

One hallmark of Etsy's shopping experience is the multitude of ways in which a user can interact with an item they are interested in: they can view it, favorite it, add it to a collection, add it to cart, or purchase it. We hypothesize that the different ways in which a user interacts with an item indicates different kinds of intent.  For example, a user who views an item must have  different intent than a user who adds the same item to their cart. Thus, the two users should be shown different recommendations, despite the fact that they both interacted with the same item. Figure~\ref{fig:interaction_target} shows an example target item, with potential recommendations for the user who \emph{viewed} that item (Figure~\ref{fig:interaction_view}) versus for the user who \emph{carted} that item (Figure~\ref{fig:interaction_cart}). Not only are the recommendations different, but the first shows recommendations that look more like substitutes to the target item, while the second shows recommendations that are more complementary. 

In this paper, we propose a novel method for learning interaction-based item embeddings that encode the co-occurrence patterns of not only the item itself, but also the way in which a user interacts with them. In contrast to previous applications of embedding models, we learn multiple embeddings for each item, one for every possible item-interaction pair.  These learned embeddings give us a convenient way of approximating the likelihood that one item-interaction pair would co-occur with another during a shopping session by way of a simple inner product. As such, we can predict not only \emph{which} items a user may be interested in, but also \emph{how} they will interact with them. Because of its computational efficiency, our model lends itself naturally as a candidate set selection process: we can generate user-specific candidate sets by finding items that lie closest to the user's past item-interaction activity in the embedded space, as visualized in Figure~\ref{figure:embedding_knn}.

Our proposed method can be seen as a generalization of using co-occurrence counts, a popular approach for candidate set selection in the past~\cite{linden2003amazon, liu2017related}. The underlying concept there assumes that if a pair of items has been viewed or purchased together within a short amount of time by the same user, there's a good chance the two items are related. However, this method (1) does not consider the different ways in which a customer can interact with items, usually focusing only on co-purchases, and (2) requires items to have been explicitly co-purchased together, leading to low coverage. Our proposed method is more flexible and generalizes beyond explicit co-occurrence counts, with the ability to give recommendations along the lines of ``Because you $X$ this, you may also want to $Y$ that'', where $X$ and $Y$ are any interaction types.

We evaluate our model as a user-specific candidate set selection method in an end-to-end production recommendation system that serves live traffic on Etsy.com. We compare our model against a live production system, which uses a co-occurrence based candidate set and provide both offline \emph{hit rate} metrics, as well as online key business metrics. Our experiments show that encoding interaction type in our item embeddings results in improved modeling accuracy. It also allows for interpretable visualization of a user's shopping behavior and can serve recommendations that offer more explainability. In the following sections, we describe related lines of work (Section~\ref{sec:related_work}), describe the proposed model (Section~\ref{sec:model}), and discuss offline and online experiment results (Section~\ref{sec:experiments}).

\section{Related Work}\label{sec:related_work}

There are two broad areas of research that closely relate to our line of work, namely (1) candidate set selection approaches and (2) the application of neural embedding models to search and ranking problems. To confine the scope of this discussion, we focus particularly on the task of recommending items to users. 

\subsection{Candidate Set Selection} 

While candidate set selection has always been an important component of large-scale recommendation systems, it is often overlooked in favor of discussing the machine learned re-ranking model. Many previous works depend on basic heuristics that measure goodness of match between the user and items based on product category or taxonomy matching schemes, or simple popularity rankings. 

Beyond basic heuristics, co-occurring signals have been a popular method for simple and efficient candidate set selection. Amazon tracks pairs of items that are frequently co-purchased by the same customers and constructs candidate sets by retrieving items that have been frequently co-purchased with a customer's last viewed items~\cite{linden2003amazon}. Pinterest's related pin recommendations system selects candidate pins based on \emph{board co-occurrence}, the number of times a pin has been pinned to the same board~\cite{liu2017related}. Other  candidate set models include variations on the random walk~\cite{eksombatchai2018pixie}, as well as collaborative filtering or deep network approaches that incorporate user's historical interactions~\cite{covington2016deep}. Another class of  algorithms come from IR, utilizing fast query-document retrieval engines to match user and item features~\cite{asadi2013effectiveness, borisyuk2016casmos}. While these methods are undoubtedly successful, our candidate set selection method explicitly considers the target item to generate candidate sets based on interaction type, whereas other methods do not distinguish between target items in a user's past history. This allows our method to be more transparent and interpretable to the user. 

\subsection{Neural Language Models}

In its original form, neural language models such as continuous bag-of-words (CBOW) and skip-gram (SG) models learn semantically meaningful, low-dimensional representation of words by modeling patterns of co-occurring words in  sentences~\cite{mikolov2013distributed}. In recent years, extending these neural embedding models to applications outside of the NLP domain has been gaining in popularity, making appearances in many domains including search, recommendations, and e-commerce~\cite{Grbovic:2018:RPU:3219819.3219885,grbovic2016ecommerce,wang2016learning,kenthapadi2017personalized}. Just as word embeddings can be trained by treating a sequence of words in a sentence as context, item embeddings can be trained in a similar fashion by treating a sequence of user actions as context and learning low-dimensional embeddings for each item. Our proposed method is very similar to this line of work; however, we explicitly model different types of interactions in addition to the item itself.

%This distinction allows us to generate candidate sets that control for the type of user interaction (e.g. purchase) that we hope to optimize for, given the specific way in which the user has interacted with (e.g. click) past items
\section{Proposed Method}\label{sec:model}

%When users come to Etsy.com, they have the option of interacting with an item in many ways. They can view or click on it; they can favorite it or add it to a collection they've created; they can also add it to their shopping cart or eventually purchase it. We hypothesize that these different kinds of interactions indicate different kinds of interest.% For example, a user who clicks on an item must have a different intent than a user who purchases the same item. As such, we focus on differentiating between item interactions so that the recommendation can be conditioned on the item, and the interaction with an item. If a user has recently favorited an item, we would want to show her items that have a high probability of purchase, given that she has favorited this item (in the vain of ``People who favorited this ended up purchasing this''). If a different user has recently purchased that same item, we may want to show a different set of items that based on the fact that the user has already purchased this item.

In the following sections, we describe a model that allows us to efficiently encode the ways in which users interact with items, and their co-occurrence relationships with other similar items. We show how to learn multiple interaction-based embeddings for each item, generate candidate sets, and produce recommmendations for users.

%In this section, we give a brief overview of Word2Vec, the neural embedding model that our proposed method is based on. We then describe our proposed model, and how it builds on top of the original skip-gram model.

%\subsection{Skip-gram Model}

\subsection{Item-Interaction Embeddings} \label{sec: ListingInteractionEmbeddings}

Our proposed model is based on word2vec, a popular method in natural language processing (NLP) for learning a semi-supervised model to discover semantic similarity across words in a corpus using an unlabelled body of text~\cite{mikolov2013distributed}. This is done by relating co-occurrence of words and relies on the assumption that words appearing together are more related than words that are far apart. 

The same method can be used to model user interactions at Etsy by modeling users' journeys in aggregate as a sequence of user engagements on listings, i.e., viewed, favored, add-to-carted or purchased. Each user session is analogous to a sentence, and each user action is analogous to a word in NLP word2vec parlance. This method of modeling interactions allows us to represent items or other entities (i.e., shops, users, queries) as low dimensional continuous vectors, where the similarity across two different vectors represents their co-relatedness. Semantic embeddings are agnostic to the content of items such as their titles, tags, descriptions, and allow us to leverage aggregate user interactions on the site to extract items that are semantically similar. 

%In this section, we describe the skipgram model we use to learn lower-dimensional representations of item interaction-type pairs. 
Let  $\Lcal$ denote the set of items and $\Ical$ the set of interaction types a user can have with an item (e.g., view, favorite, add-to-cart, purchase). Each user's visit on Etsy triggers a \textit{session} $S=\{p_1,\dots,p_k\}$, which consists of a sequence of item-interaction pairs $p_j \in \Lcal \times \Ical$.
For example, the sequence $$(\ell_1,\text{view}),  (\ell_2,\text{favorite}), (\ell_1,\text{purchase})$$ specifies that a user first viewed item $\ell_1$, then favorited item $\ell_2$, and lastly purchased item $\ell_1$.  The training data consists of such sequences collected from multiple users over a set period of time. 

The skip-gram model uses a single item-interaction pair to predict the output of $2m$ neighboring pairs, where $m$ is a hyperparameter of the model. 
% The context contains $m$ item-interaction pairs corresponding to $m/2$ pairs both before and after the \textit{target} item-interaction pair.\footnote{We assume, without loss of generality, that $m$ is an even number.}
In particular, given an item-interaction pair $p_i \in \Lcal \times \Ical$, the probability of observing the pair $p_{i+j}$ is given by 
\begin{align}\label{eq:prob}
	\Prob(p_{i+j} \ | \ p_i ) = \frac{u_{p_i}^\top v_{p_{i+j}}}{\displaystyle \sum_{p=1}^n u_{p_i}^\top v_{p} }, \quad -m \le j \le m,
\end{align}
where $u_{p}, v_{p} \in \R^d$ are the input and output vector representations of the pair $p$, repsectively and $n$ is the number of unique item-interaction pairs. Implicit in the skip-gram model is the assumption that the dimension $d << n$. The objective is to maximize the function 
\begin{align}\label{eq:objective_function}
     \sum_{S \in \Scal} \sum_{p_i \in S} \sum_{-m \le j \le m} \Prob(p_{i+j} \ | \ p_i ),
\end{align}
where $\Scal$ denotes the set of all sessions in the training data. It follows from \eqref{eq:prob} and \eqref{eq:objective_function} that item-interaction pairs with similar contexts will have ``similar'' vector representations. The objective function is optimized using stochastic gradient ascent.
In practice, however, the computation of an optimal solution, can be computationally expensive because the size of the ambient dimension space $|\Lcal| \times |\Ical|$ can be prohibitively large. To account for this, we use the negative sampling approach proposed in \cite{mikolov2013distributed}. We provide more details for the approach we take to generate negative samples in section \ref{sec:negative_sampling}.

\subsection{Candidate Set Selection}

A crucial task for industry-scale recommendation systems is its ability to quickly retrieve a small set of relevant items out of a large set (potentially in the range of hundreds of millions) of candidate items. This is necessary as it is computationally infeasible to apply machine-learned models over the entire collection of candidate items. In practice, this is referred to as the \emph{candidate set selection} phase, which aims to quickly prune irrelevant items while retrieving items that are likely to be relevant to the user at low cost. The smaller set is then re-ranked by a (typically, more sophisticated) machine learning model.

\begin{figure}[t]
\includegraphics[scale=0.3]{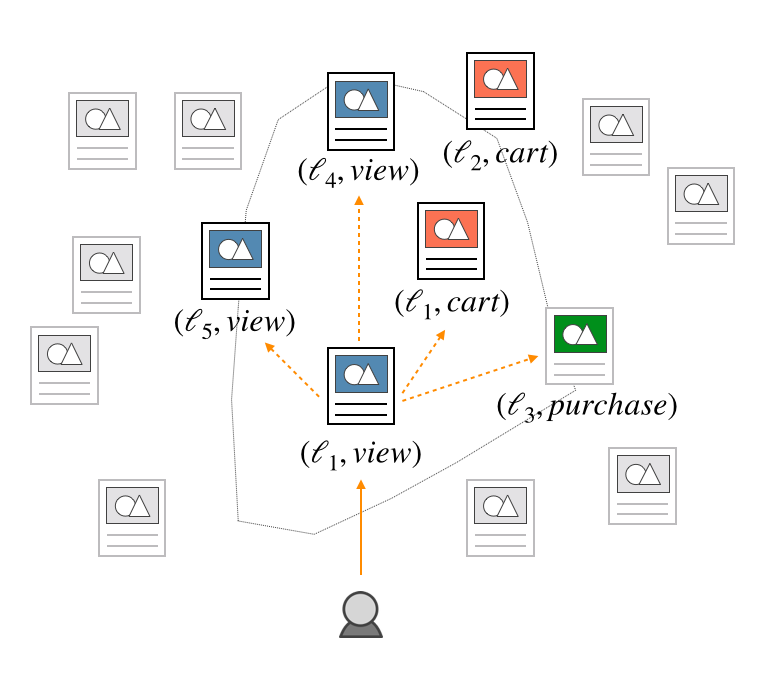}
\caption{Example of the item-interaction pair embedding space. Given a user's last action (e.g. viewed listing $\ell_1$), the top $K$ nearest neighbors should indicate \emph{which} item the user is likely to interact with, as well as \emph{how} the user will interact with it (e.g. the user may \emph{view} items $\ell_4$ or $\ell_5$; \emph{cart} items $\ell_1$ or $\ell_2$; or \emph{purchase} item $\ell_3$) }
\label{figure:embedding_knn}
\end{figure}

The learned embeddings described above now give us a convenient way of encoding co-occurrence patterns between items and the way users interact with them. A nice property is that the inner product between two such embeddings should approximate the likelihood that one item-interaction pair would co-occur with another. Because of its computational efficiency, it is easy to approximate the affinity between tens of millions of pairs of embedding vectors, lending itself naturally as a candidate set selection solution. For example, to answer a question such as ``since a user viewed on item A, what is an item they may add to cart next?'', we can simply find the nearest ``cart'' embedding to item A's ``view'' embedding. A user-specific candidate set can then be generated by finding the closest listings to each of the user's past item-interaction pairs in the embedded space. Figure~\ref{figure:embedding_knn} explains this idea visually.

\section{Experiments} \label{sec:experiments}
In this section, we investigate the strengths and limitations of the proposed
approach and discuss evaluation results on a dataset of browsing sessions from Etsy.

\subsection{Dataset} \label{sec:Exp_dataset}
The training data we use spans a one year period of visit log collected from Nov 2017 to Oct 2018, and is extracted from implicit feedback collected from users visiting Etsy during that period. In particular, each training instance is defined by a user's session and consisted of a sequence of item-interaction pairs sorted chronologically.  We restrict attention to sessions which had more than three item-interaction pairs to eliminate bounces. The resulting dataset has about 30 billion words from over 200 millions distinct tokens.  

\subsubsection{Negative Sampling} \label{sec:negative_sampling}
As discussed in section \ref{sec: ListingInteractionEmbeddings} the model we propose uses the negative sampling approach introduced in \cite{mikolov2013distributed} to facilitate computational efficiency in model training and improve the quality of the low-dimensional vector representations. As the initial step in the implementation of negative sampling, we define the following partial order on the set of interactions we consider:
\[
    \text{purchase} > \text{add-to-cart} > \text{favorite} > \text{view}.
\]
For each item-interaction pair $p = (\ell, i)$ of a given session $S$, we add negative samples $p_j= (\ell, j)$, for all interactions $j > i$, provided that the pair $p_j \not \in S$. For example, if an item has been viewed and added-to-cart, we associate with it a \textit{negative purchase}. A natural drawback of this approach is that it precludes the addition of negative samples having \textit{view} as an interaction. To account for this, we include two negative pairs $(\ell_1, \text{view}), (\ell_2, \text{view})$, for each item $\ell$ that was only viewed. The items $\ell_1,\ell_2$ are drawn uniformly at random from the set of items belonging to the same taxonomy as $\ell$ in order to capture the user's preference in the item viewed in said taxonomy.

\begin{figure}[t]
    \centering
\hspace{-0.73cm}    \begin{overpic}[scale=0.6]{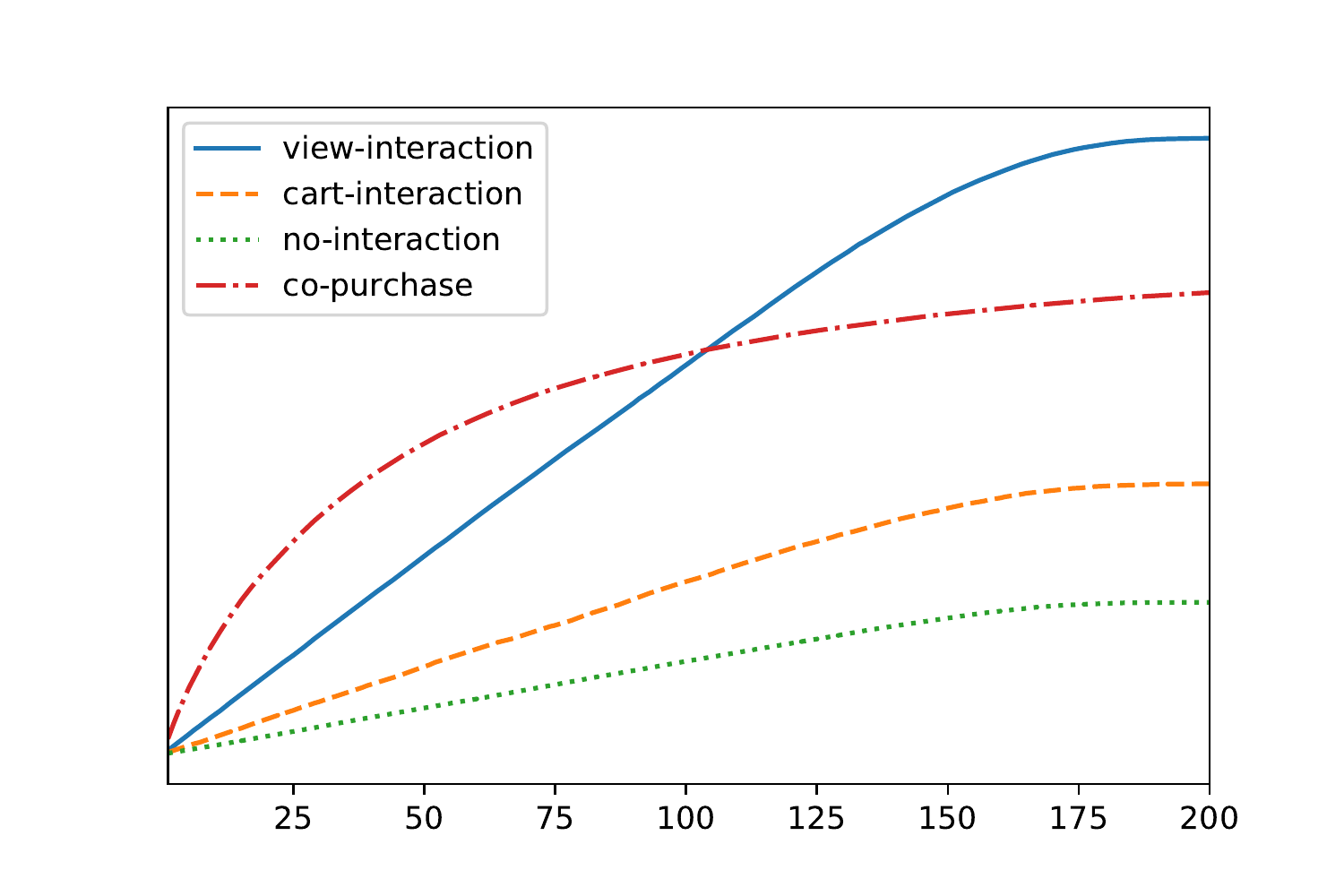}
    \put (40,0.0) { \normalfont{\# of candidates}}
    \put (6,29.5) { \rotatebox{90}{\normalfont{$h^{\text{purchase}}$}} }
    \end{overpic}
    \caption{Comparison of average hit rates among different candidate sets with the set of purchased items.}
    \label{fig:avg_hit_rates}
\end{figure}

\subsection{Implementation Details}
In order to support custom negative sampling, we used the \textit{fastText} library \cite{bojanowski2016enriching} developed by Facebook's AI research lab, to train our embedding models with extended functionalities built on top of the existing library. The primary innovation from \textit{fastText} is that it enriches the word2vec model \cite{mikolov2013distributed} with subword information by adding a bag of character ngrams. We chose the framework in favor of its easiness in extensibility as well as its efficiency in training. 
% Various hyperparameters are tuned in training, and we then finalized to set co-occurence neightborhood with a window size $m=5$, embedding dimension of $d=100$, and random negative sampling of 5 in addition to the above mentioned custom negatives.  
We experimented with tunning several hyperparameters of the model and eventually chose to set the context window to $m=5$ and the embedding dimension to $d=100$ (cf. section \ref{sec: ListingInteractionEmbeddings}). In addition to the aforementioned custom negative samples, we also added five random negative samples in each of our sequences.

After training, we use the approximate K-nearest neighbor search algorithm, \textit{Hierarchical Navigable Small World} \cite{malkov2016} from Faiss (HNSW) \cite{JDH17}, in order to get the top $k$ similar items for each learned item-interaction pair. To balance efficiency with accuracy, we set the following hyperparameters in HNSW  ($\emph{efSearch}=256$, $\emph{efConstruction}=128$, $\emph{linksPerVector}=64$) to scale up the approximated neighborhood search over hundreds of millions item-interaction pairs from our embedding training.

\subsection{Offline Experiments}
We test the performance of our model on data collected from shopping sessions that start in the 24 hour window following the last day of training. 

\subsubsection{Evaluation Methodology}
% TODO: emphasize that this is for candidate set selection portion only (and explain why we don't use AUC on historical data)

We evaluate our candidate set selection method by computing a \textit{hit rate}, a metric similar to recall for the candidate sets we generate. (We use hit rate as a metric instead of the the more commonly-used AUC because the candidates we generate are not necessarily present in historical data.) To do so, we first fix a target interaction $i$, and an item-interaction pair $p$ and define the set 
\begin{align*}
    \Hcal_{p}^i \coloneqq \left\{ \left(\ell, q\right) \ \Big| \ p \neq (\ell, i) \in S, \ q = \sum_{S \in \Scal} q_{S}(\ell,i) \right\},
\end{align*}
where $q_{S}(\ell,i) \in \{1,2,\dots\}$ denotes the number of times item $\ell$ had interaction $i$ in session $S$. For example, if $p=(\ell,\text{view})$ and $i$= purchase, then $\Hcal_{p}^{i}$ contains all items (with associated quantities) purchased across all sessions after item $\ell$ was viewed.
The hit rate of the $p^{\rm th}$ item-interaction pair is given by
\begin{align*}
   h_p^i \coloneqq \displaystyle \sum_{j=1}^{|\Hcal^i_p|} q_j \bone_{\{\ell_j \in \Ccal_p\}} \Biggm/ \displaystyle \sum_{j=1}^{|\Hcal^i_p|} q_j,
\end{align*}
%\begin{align*}
%   h_p^i \coloneqq \frac{\displaystyle %\sum_{j=1}^{|\Hcal^i_p|} q_j \bone_{\{\ell_j %\in \Ccal_p\}}}{ \displaystyle %\sum_{j=1}^{|\Hcal^i_p|} q_j},
%\end{align*}
where, $\Ccal_p$ denotes the candidate set of the $p^{\rm th}$ pair produced by our model and $\bone$ is the indicator function.
The average hit rate across all item-interaction pairs for a target interaction $i$ is given by
\begin{align} \label{eq:avgHitRate}
   h^i \coloneqq \frac{1}{k}  \sum_{p \in S} h_p^i,
\end{align}
where $k$ denotes the number of disctinct item-interaction pairs in the test set.

% More precisely, fix a session $S$, a target interaction $i$, and a pair $p$ and define the set 
% \begin{align*}
%     \Hcal_{p}^i(S) \coloneqq \Big\{ (\ell_j, q_j) \ \Big| \ (\ell_j, i) \in S, \ (\ell_j, i) \neq p \Big\},
% \end{align*}
% where $q_j \in \{1,2,\dots\}$ denotes the number of times item $\ell_j$ had interaction $i$ in session $S$. 
% For example, if $p=(\ell,\text{view})$ and $i$= purchase, then $\Hcal_{p}^{\rm i}(S)$ contains all items (with associated quantities) purchased in session $S$ after item $\ell$ was viewed.
% % As a first step in evaluating our model we compute the hit rate for each item-interaction pair on our test set, where 
% The hit rate of the $p^{\rm th}$ pair is given by
% \begin{align*}
%   h_p^i(S) \coloneqq \frac{\displaystyle \sum_{j=1}^{|\Hcal_i|} q_j \bone_{\{\ell_j \in \Ccal_p\}}}{ \displaystyle \sum_{j=1}^{|\Hcal_i|} q_j},
% \end{align*}
% where, $\Ccal_p$ denotes the candidate set of the $p^{\rm th}$ pair produced by our model and $\bone$ is the indicator function.
% The average hit rate across all item-interaction pairs for a target interaction $i$ is given by
% \begin{align*}
%   h^i \coloneqq \frac{1}{\displaystyle \sum_{S\in \Scal} |S|} \sum_{S \in \Scal } \sum_{p \in S} h_p^i(S),
% \end{align*}
% where $|S|$ is equal to the number of disctinct item-interaction pairs in $S$.

\subsubsection{Experimental Results}
For offline experiments, we choose two main models for candidate set generation for comparison: \textbf{View-Interaction}, which generates a candidate set based only on the ``view-based'' embeddings for a listing, and  \textbf{Cart-Interaction}, based only on ``cart-based'' embeddings for a listing. We compare these methods with the following two baselines: 
\begin{itemize}
\item{\textbf{Co-Purchase:}} This method represents each item as a \emph{co-purchase vector}, a sparse vector indicating other items that have been co-purchased with it, similar to~\cite{linden2003amazon}. The candidate set is generated by finding the top $K$ items with the most similar co-purchase vectors as the target item. We chose this as a baseline because it is the dominant candidate set selection method currently used in Etsy's production systems.
\item{\textbf{No-Interaction:}} This method learns a single embedding for each item without differentiating between interaction types, similar to~\cite{Grbovic:2018:RPU:3219819.3219885}. The candidate set is also generated by finding the top $K$ items with the most similar embedding to the target item. 
\end{itemize}

\begin{table}[t]
\setlength{\tabcolsep}{7.3pt}
\renewcommand{\arraystretch}{1.0} % General space between rows (1 standard)
\begin{tabular*}{3.3in}{ccccccc} 
\toprule[1.2pt]\midrule[0.3pt]
 & & \multicolumn{5}{c}{Test Target Interaction}\\[0.5em] 
\multirow{2}{*}{Model} & & View & & Cart & & Purchase \\[-0.2em]
 & & (\%) & & (\%) & & (\%) \\[0.2em]
  \cmidrule{3-3} \cmidrule{5-5} \cmidrule{7-7} 
View-interaction & &  201.90 & &  198.11  & & 207.10   \\
Cart-interaction  & &   -6.88 & &  34.88 & & 47.74  \\                 
Co-purchase  & &   76.89 & &  179.70 & &  129.60 \\        \midrule      
\end{tabular*}
\caption{Percent change in hit rate compared to the model with no-interactions for embedding vectors of dimension $d =100$ and candidate sets of size 200.} 
\label{table:Percent_change_in_hit_rates}
\end{table}

In Figure \ref{fig:avg_hit_rates}, we plot the average hit rate defined in \eqref{eq:avgHitRate} for all candidate set methods as a function of the number of candidates in the set number of items in them
when the target interaction is a purchase.
% the average hit rate among different candidate sets when the target iteraction is a purchase.
% In Figure \ref{fig:avg_hit_rates}, we compare the average hit rate among different candidate sets when the target iteraction is a purchase.
We observe that the candidate set associated with the view-interaction model outperforms all other candidate sets, including the co-purchase baseline when the number of items it contains is sufficiently large.
Although the view-interaction model underperforms the co-purchase model for candidate sets with smaller number of candidates, it is beneficial to explore additional items in candidate selection as this can increase the average hit rate.
% eventually leads to higher hit rate.
%The co-occurrence approach based on the wisdom of crowd or memorization on historical counts provides great initial warm-start, but generation learned from the item-interaction model quickly outpaces it as sizes increase. The next layer, re-ranker, optimizes toward main objectives given the sizable set of relevant items.         
Additionally, we observe in Figure \ref{fig:avg_hit_rates} that the no-interaction model has the smallest average hit rate among all other models. Therefore, learning an embedding for item-interaction pairs instead of a single embedding for each item has resulted in a candidate set which contained more items that were eventually purchased.

In Table \ref{table:Percent_change_in_hit_rates}, we observe that the view-interaction model outperforms the no-interaction model as well as all other models for all target interactions we consider. The cart-interaction model is shown to outperform the no-interaction model only when the test target interaction is either a cart or purchase action. However, it is not shown to outperform the co-purchase baseline. We believe that this is due to sparsity of add-to-cart interactions
in our training data, resulting in suboptimal embedding representations. 

\begin{table}[t]
\setlength{\tabcolsep}{10.0pt}
\renewcommand{\arraystretch}{1.0} % General space between rows (1 standard)
\begin{tabular*}{3.3in}{ccccccc} 
\toprule[1.2pt]\midrule[0.3pt]
\multirow{4}{*}{Model} & & \multicolumn{4}{c}{Coverage Rate ($\%$ of active listings)}\\[0.4em]
 & &  Items & & Traffic \\[-0.2em]
 & & (\%) & & (\%)  \\[0.2em]
  \cmidrule{3-3}  \cmidrule{5-5} 
View-interaction & &   73.98  & & 85.34  \\
Cart-interaction  & &  70.57 & & 80.13   \\ 
No-interaction  & &  78.51 & & 96.11 \\   
Co-purchase  & & 9.43 & &  42.83  \\ \bottomrule                 
\end{tabular*}
\caption{Target item coverage rate, based on distinct active items and percent of traffic on one-day visit logs.}
\label{table3:coverage_rate}
\end{table}

%% Raphael: Changed to this, but feel free to change back/modify
One disadvantage of relying on historical co-interactions (i.e., co-purchases) for candidate set selection is their relatively low coverage rate.
This is evident in Table \ref{table3:coverage_rate}, which shows that the existing \emph{co-purchase} baseline covers
only 9.43\% of active items. This is due to stringent requirements for constructing such candidate sets. For example, the co-purchase candidate set requires two or more items to be purchased within a small time window.
Such critieria is hardly applicable for the majority of items at Etsy due to the one-of-a-kind nature for the majority of the items. In addition, low coverage is also evident in the size of the candidate set for the co-purchase based candidate set (approximately $40$ items).
As shown in Table \ref{table3:coverage_rate}, both the item-interaction and no-interaction methods cover at least 70\% of distinct active items, which account for more than 80\% of %item engagement on 
Etsy's traffic.
% One disadvantage of using historical co-occurrence (i.e., copurchase) to generate candidate set is its lower coverage rate, as shown in Table \ref{table3:coverage_rate}.
% Only 9.43\% of active items have candidate sets using our existing \emph{co-purchase} approach, which requires two items to be co-purchased within some time window. This critieria is hardly applicable for the majority of items at Etsy due to its one-of-a-kind nature; this is again evident in the lower average cardinality in candidate set size, (approximately $40$ items). As shown in Table \ref{table3:coverage_rate}, both the item-interaction and no-interaction approaches cover at least 70\% of distinct active items, which accounts for more than 80\% of %item engagement on 
% Etsy`s traffic.

\begin{table}[t]
\setlength{\tabcolsep}{7.3pt}
\renewcommand{\arraystretch}{1.1} % General space between rows (1 standard)
\begin{tabular*}{3.3in}{ccccccccc} 
\toprule[1.2pt]\midrule[0.3pt]
 & & \multicolumn{4}{c}{Embedding Dimension}\\[0.5em]
\multirow{2}{*}{Model} & & $d=25$ & & $d=50$ & & $d=75$ \\[-0.2em]
 & & (\%) & & (\%) & & (\%) \\[0.2em]
  \cmidrule{3-3}  \cmidrule{5-5} \cmidrule{7-7} 
View-interaction & &  -7.51   & & -1.31 & & -1.85  \\
No-interaction  & &  -12.95 & & -0.19 & &  -0.16 \\               \bottomrule                 
\end{tabular*}
\caption{Percent change in hit rate for item-interaction embedding models of different dimension compared to the model using $d=100$.
The test interaction used is purchase and the candidate set size is equal to 200.}
\label{table2:percent_change_by_dimension}
\end{table}

Table \ref{table2:percent_change_by_dimension} shows performance metrics for the item-interaction and no-interaction models as a function of the embedding dimension. The numbers reported are average hit-rates for candidate sets with 200 items.
% Table \ref{table2:percent_change_by_dimension} shows how the item-interaction and no-interaction models perform when the embedding dimension varies.
For each embedding dimension, $d \in \{25, 50, 75\}$, the table shows its percent change in hit rate compared to its corresponding model type with $d=100$. In both model types, we observe a drastic deterioration in performance for smaller dimension, $d=25$. With Etsy's many one-of-a-kind listings, $d=25$ is likely not large enough to capture the variance. As a result, we see the hit rate for both models improve as $d$ increases. However, this marginal gain begins to saturate as $d > 100$ as computation feasibility begins to decrease model performance. Offline, we observe an almost linear trend in training time and machine memory in dimension $d$.

\subsection{Online Experiments}

\begin{figure}[t]
\begin{subfigure}{0.515\textwidth}
  \includegraphics[width=.9\linewidth]{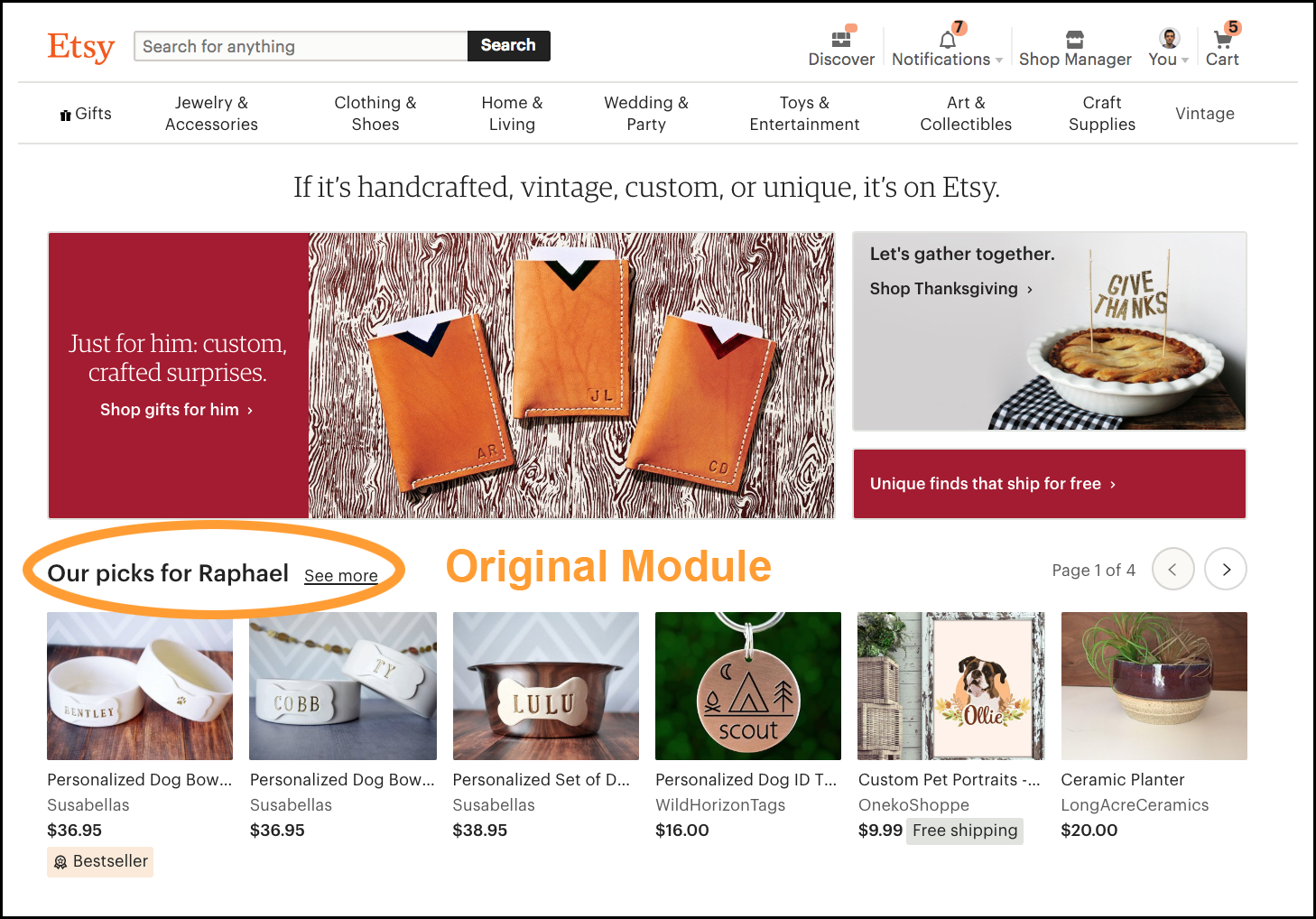}
   \caption{Control Group}
  \label{fig:sub1}
\end{subfigure}\\[0.5em]
\begin{subfigure}{0.515\textwidth}
  \includegraphics[width=.9\linewidth]{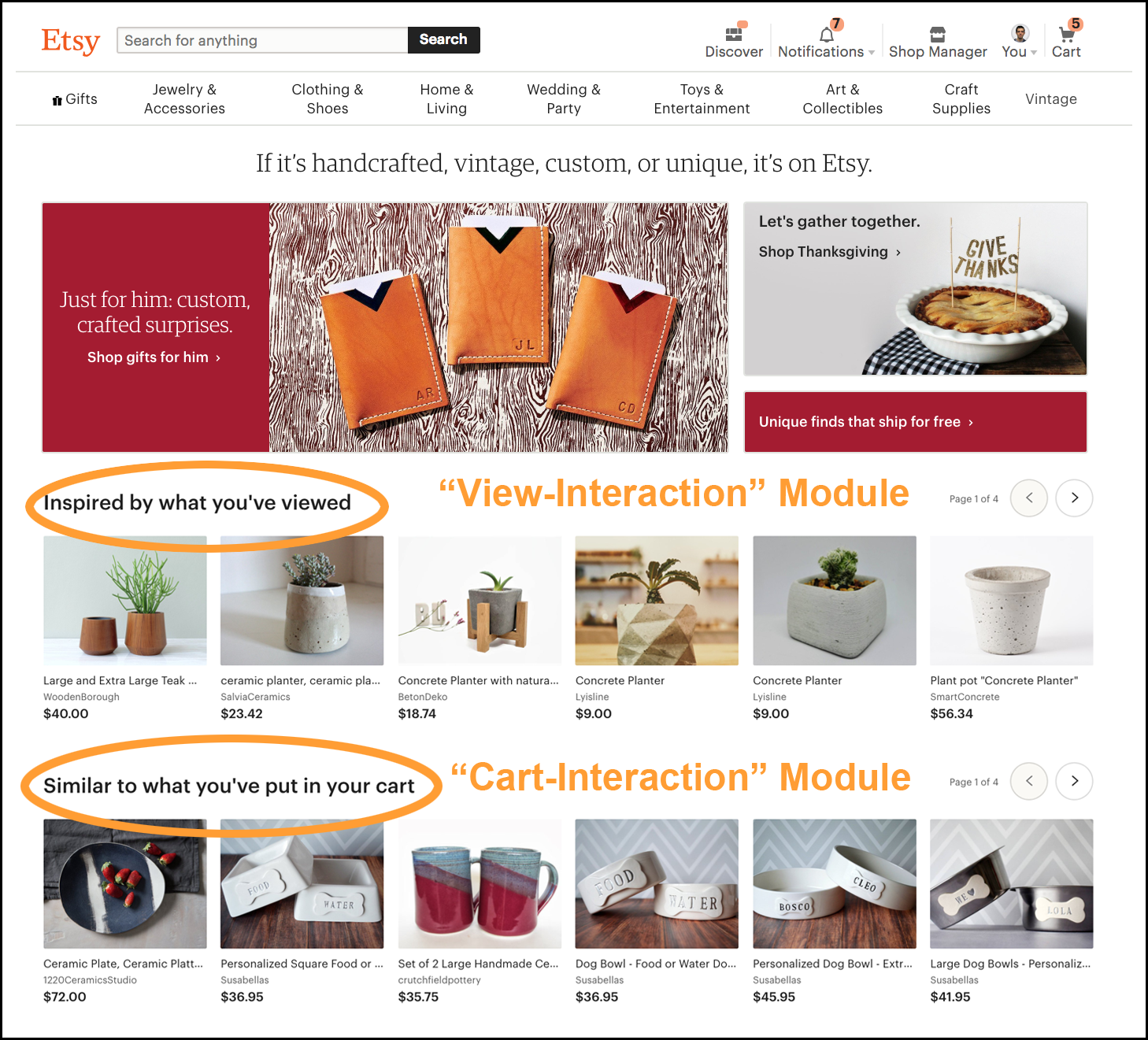}
    \caption{Treatment Group}
  \label{fig:sub2}
\end{subfigure}
\caption{Visual examples of the recommendations received by users in the two buckets of our live A/B test, including: (a) the control group, which used a co-purchase based candidate set, and (b) the treatment group, which used the proposed item-interaction embedding model. The single recommendation module was split into two separate modules based on the way the user interacted with past listings. This was preferred as it offers more explainability to the user.}
\label{fig:test}
\end{figure}

To further evaluate our candidate set selection methodology, we implement and deploy the algorithm to a production user recommendation module on the homepage of Etsy.com. We train an embedding model based on Section~\ref{sec:negative_sampling} and generate candidate sets for each user as described below. 

A live A/B test was run for 7 days in which we bucketed $50\%$ of signed-in users into the \emph{control} group, and the remainder into the \emph{treatment} group. The \emph{control group} received a single module that recommends items using a candidate set that matched items based on the last $100$ items that the user interacted with, regardless of the interaction type. The \emph{treatment group} received two separate modules: The first module finds the closest item-interaction embeddings based on the last 4 items that the user had viewed; the second module is computed similarly but is based on the last 4 items that the user has added to cart. Figure~\ref{fig:test} shows a visual description of what the two variants look like.

Both the control and treatment groups used the same re-ranking model on top of the candidate set selection phase, which has an objective function that is optimized for predicting purchases, and uses a variety of historical user and listing rates and content features. We also note that while we prepared additional modules based on favoriting and purchasing behavior, they were not deployed to the live A/B test due to their lower-than-desired offline metrics.  

The online A/B test evaluates an end-to-end system, which includes a combination of several factors, not only the choice of the candidate set selection algorithm. These additional factors include: the design for  breaking up the original module into two separate modules, the change of copy on the new modules, as well as the effect of the re-ranker on top of the candidate set. As such, we look at high-level business metrics to evaluate an end-to-end system with potentially many confounding factors. 

For this particular experiment, the metrics that we tracked include site-wide click-through-rate, conversion rate, and add-to-cart rate. Given the preliminary experimental data, we observed promising upward trend in many key metrics, although more data is needed to gain statistical significance. Compared to the \emph{control}, the first module in the treatment (``Inspired by what you've viewed'') showed 4.1\% improvement in click-through-rate. Additionally, the ``treatment'' group showed 0.20\% and 0.31\% increase in conversion rate and checkout rate, respectively.

%\begin{figure}
%\centering
%\subfloat[Recommendations received by the %control group]{
%  \includegraphics[scale=0.20]{images/Control.png}
%}
%\hspace{0mm}
%\centering
%\subfloat[Recommendations received by the treatment group]{
  %\includegraphics[scale=0.20]{images/Treatment.png}
%}
%\caption{Without giving away Etsy, show something that will easily illustrate what a recommendation module is and how users interact with it and how we collect click and purchase data from it. 
%}
%\label{figure:recommendation_overview}
%\end{figure}

\section{Conclusion}

In this paper, we describe a model for learning interaction-based item embeddings, which we use for candidate set selection. The proposed method allows us to encode co-occurance patterns between items and the way users interact with them. We train our model on a large production dataset of browsing sessions from Etsy and evaluate its performance both through offline metrics and online experiments. We observe that the canidate sets produced by our model improve upon the current production baselines.

\bibliographystyle{ACM-Reference-Format}
\bibliography{acmart.bib}
% \bibliography{sample-bibliography}

\end{document}